# Programmable Phase-change Metasurfaces on Waveguides for Multimode Photonic Convolutional Neural Network

Changming Wu[1], Heshan Yu[2], Seokhyeong Lee[1], Ruoming Peng[1], Ichiro Takeuchi[2],

and Mo Li[1,3]*

[1]Department of Electrical and Computer Engineering, University of Washington, Seattle, WA 98195, USA

[2]Department of Materials Science and Engineering, University of Maryland, College Park, MD 20742, USA

[3]Department of Physics, University of Washington, Seattle, WA 98195, USA

**ABSTRACT**

**Neuromorphic photonics has recently emerged as a promising hardware accelerator, with significant potential speed and energy advantages over digital electronics for machine learning algorithms such as neural networks of various types. Integrated photonic networks are particularly powerful in performing analog computing of matrix-vector multiplication (MVM) as they afford unparalleled speed and bandwidth density for data transmission. Incorporating nonvolatile phase-change materials in integrated photonic devices enables indispensable programming and in-memory computing capabilities for on-chip optical computing. Here, we demonstrate a multimode photonic computing core consisting of an array of programable mode converters based on on-waveguide metasurfaces made of phase-change materials. The programmable converters utilize the refractive index change of the phase-change material $Ge_2Sb_2Te_5$ during phase transition to control the waveguide spatial modes with a very high precision of up 64 levels in modal contrast. This contrast is used to represent the matrix elements, with 6-bit resolution and both positive and negative values, to perform MVM computation in neural network algorithms. We demonstrate a prototypical**

* Corresponding author: moli96@uw.edu

**optical convolutional neural network that can perform image processing and classification tasks with a high accuracy. With a broad operation bandwidth and a compact device footprint, the demonstrated multimode photonic core is readily scalable toward large-scale photonic neural networks with ultrahigh computation throughputs.**

The unmet gap between the rate of energy efficiency improvement of current digital electronics and the fast-growing load of computation by emerging applications such as machine learning and artificial intelligence[1,2] has once again brought optical computing into focus[3-6]. Integrated photonics provides a scalable hardware platform to realize large-scale optical networks on a chip, which affords an enormous bandwidth density that is unreachable for electronics[7-9]. To use integrated photonics for optical computing, programmable photonic components and nonlinear elements are indispensable building blocks. Phase-change materials (PCM) recently emerged as an ideal material system to realize optical programmability[10-12]. The optical properties of PCMs change dramatically during the phase transition, which can be electrically or optically controlled. Harnessing this has allowed for embodiments of programmable optical switches, couplers, lens and metamaterials to be demonstrated[13-21]. The phase change in the chalcogenide family of Ge-Sb-Te alloys is nonvolatile, requiring no sustaining power supply to retain the programmed state or stored information[19-26]. Their use in programmable photonic devices thus can have a significant advantage in power consumption over electro-optic[27-29] or thermo-optic methods[30-32]. Photonic devices incorporating those nonvolatile PCMs thus can realize optical memories and perform in-memory computing simply by measuring the transmission of the optical input data through the programmed device[33-35]. Proliferating these phase-change photonic devices in a scalable network, prototypes of an optical neural network (ONN), has been proposed and demonstrated[35-38].

Here, we report a programmable waveguide mode converter based on a phase-gradient metasurface made of phase-change material $Ge_2Sb_2Te_5$ (GST). This phase-change metasurface mode converter (PMMC) utilizes GST's large refractive index change during its phase transition to control the conversion of the waveguide's two spatial modes ($TE_0$ and $TE_1$ modes). The PMMC can be programmed to control the waveguide mode contrast precisely at 64 distinguishable levels, which is used to represent the weight parameters with 6-bit precision in MVM computation. We build a 2×2 array of PMMCs and implement them as programmable kernels to realize a multimode optical convolutional neural network (OCNN). By performing image processing tasks such as edge detection and pattern recognition, we demonstrate the OCNN's viability and potential in large-scale optical computing.

The design of the PMMC is based on the principle of a phase-gradient metasurface but replacing noble metals with phase-change materials[39]. Fig. 1a shows a 3D schematic of the design, which consists of a linear array of GST nano-antennae directly integrated on a silicon nitride (SiN) waveguide. Each GST nano-antenna scatters the waveguide mode and causes a phase shift $\Phi$, which depends on its geometry (e.g., width), as well as the refractive index of its material (Fig. 2b). A linear array of such nano-antennae with tapering widths thus produces a spatial gradient of the scattering phases $d\Phi/dx$, which is equivalent to a wavevector $k_g$. If the phase-gradient metasurface is designed such that $k_g$ matches the wavevector difference between two spatial modes of the waveguide: $k_{mode1}-k_{mode2}$, it satisfies the phase-matching condition and facilitates the conversion between the two modes. Such phase-gradient metasurfaces for waveguide mode conversion realized with noble metals or dielectrics materials thus lacked tunability. Here, we use GST, which has a large change in its optical properties when a phase transition happens. When the GST is in the amorphous phase (aGST), its refractive index $n$ is ~4.7 (representative value in the literature, the same hereafter)[40]. In contrast, when it is turned to the crystalline phase (cGST), $n$ increases to ~7.5 with a drastic change of 2.8 over the whole measured spectral range from 1540 nm to 1580 nm (See Figure S1a for more detailed information). This change will significantly modify the scattered phase of each GST nano-antenna (Fig. 2b) so as to modify the metasurface's function. Fig. 1c plots the simulated phase of the scattered fields inside the waveguide by a single nano-antenna of 30-nm-thick GST as a function of its width and for aGST and cGST phases. Since cGST has a much larger $n$, the scattered phase shows a much stronger dependence on the width than the aGST phase. By controlling the geometry of the GST nano-antennae and the interval between adjacent ones in the array, a well-defined phase gradient $d\Phi/dx$ is established (See Supplementary Information for details). The entire metasurface consists of an array of 25 nano-antennae with tapering widths from 510 nm to 84 nm (shaded region in Fig. 1c) and is patterned on a SiN waveguide 1.8 μm wide and 330 nm thick. The waveguide supports two transverse-electric modes: the fundamental $TE_0$ mode and the first-order $TE_1$ mode. We design the metasurface, in the cGST phase, to have a uniform $d\Phi = 2.5°$ for every $dx = 400$ nm to satisfy the generalized phase-matching condition, $k_0(n_{TE0} - n_{TE1}) = N \cdot d\Phi/dx$, where $k_0$ is the free-space wavevector, $n_{TE0}$ and $n_{TE1}$ are the effective index of the TE$_0$ and TE$_1$ modes, respectively, and $N$ is the number of interactions between the guided modes and the metasurface. The cGST metasurface thus can efficiently convert the TE$_0$ mode to the TE$_1$ mode, as shown by the finite-domain finite-time (FDTD) simulation result in Fig. 1d. When the GST is transitioned to the aGST phase, as shown in Fig. 1c, the $d\Phi/dx$ is much reduced and thus insufficient for the phase-matching condition so that mode conversion between TE$_0$ and TE$_1$ modes does not occur, which is clearly seen in Fig. 1e. Therefore, the GST phase-gradient metasurface, as designed here, functions as a programmable waveguide mode converter controlled by the tunable material phase of the GST.

Fig. 2a-c shows the scanning electron microscope images of the complete PMMC device. The 30nm thick GST film is deposited by sputtering on Si$_3$N$_4$ on an oxided silicon substrate. It is then patterned into metasurface with electron beam lithography and plasma etching, and conformally encapsulated with a 218 nm thick layer of Al$_2$O$_3$ deposited by atomic layer deposition. The photonic circuits of Si$_3$N$_4$, including multimode waveguides, directional couplers and grating couplers, are then patterned with standard processes[17]. A pair of asymmetric directional couplers (Fig. 2c) is designed to function as mode selectors to selectively couple only the TE$_1$ mode component in the multimode waveguide with the TE$_0$ mode component in the single-mode waveguide (See Supplementary Information for details). Fig. 2a depicts the measurement and control scheme. To program the PMMC, we use optical pulses to control the phase of the GST film for simplicity[41]. Previously, electrical control using integrated micro-heaters has been previously demonstrated by a number of groups, including us[17,26,42-44]. When operating the PMMC, an optical signal is input in the TE$_0$ mode to the PMMC and converted to TE$_1$ mode with a proportion controlled by the state of the GST metasurface. At the output of the PMMC, the TE$_1$ component is separated by the mode selector and coupled out at the second port while the TE$_0$ component remains in and outputs from the multimode waveguide. The output powers of both modes are then measured to determine their respective transmission coefficients. Fig. 2d shows the transmission spectrum of the PMMC when the metasuface is set to be either in fully aGST or cGST phases. The insertion losses of the input and output fibers and grating couplers have been accounted for by calibration measurements. In the aGST phase, the device is in the on-state for the TE$_0$ mode with a high transmission $T^{on}$ over a broad wavelength range (1540 to 1580 nm). The lowest insertion loss is 0.9 dB at 1575 nm wavelength. A small portion (<-10 dB) of the TE$_1$ mode is generated due to the asymmetric perturbation induced by the metasurface even though the aGST

phase has a low refractive index. The situation changes dramatically when the metasurface is transitioned to the cGST phase and converts the TE$_0$ mode to the TE$_1$ mode effectively. In this off-state for the TE$_0$ mode, its transmission $T^{\text{off}}$ is lower than -15 dB over the entire measured bandwidth. The corresponding switching extinction ratio, defined as $\Delta T/T^{\text{off}} = (T^{\text{on}} - T^{\text{off}})/T^{\text{off}}$, is ~16 dB or 4000%, which is more than 10-fold improvement compared to previously reported switch devices using GST [24,43,45]. This large switching ratio stems from the phase engineering approach to effectively use GST's large refractive index change during its phase-transition, as opposed to only using the absorption coefficient change, to facilitate scattering into a different mode that is filtered. The total area of the GST in the metasurface is only 1.3 μm$^2$, significantly smaller than that in prior devices, and thus in principle, our device consumes less energy to switch. As expected from energy conservation, the TE$_1$ mode is switched in the opposite way to the TE$_0$ mode. From aGST to cGST phase, the TE$_1$ transmission increases from ~-10 dB to ~-6.5 dB, with the insertion loss due to cGST's absorption. Another important parameter to quantify a mode converter's performance is the mode purity in the multimode waveguide, defined as $\beta_{\text{TE0(TE1)}} = P_{\text{TE0(TE1)}}/(P_{\text{TE0}} + P_{\text{TE1}})$, where $P_{\text{TE0}}$ ($P_{\text{TE1}}$) is the power in the TE$_0$ (TE$_1$) mode. The PMMC shows very high performance in controlling mode purity. As shown in Fig. 2e, when switching the GST from aGST to cGST phase, the PMMC efficiently converts TE$_0$ mode to TE$_1$ mode, changing the mode purity from $\beta_{\text{TE0}}$>80% to $\beta_{\text{TE1}}$>85% over a broad bandwidth, showing an excellent agreement with the numerical simulation results.

The phase composition of the GST in the metasurface can be continuously tuned by partial phase transition so that the PMMC can be continuously programmed to multiple intermediate levels of phase purity values. We program the PMMC with a sequence of 50 ns-long control pulses to "quench" the GST progressively from the fully cGST phase toward the fully aGST phase. As a result, the TE$_1$ mode purity $\beta_{\text{TE1}}$ increases stepwise. Since the mode selector separates the two modes, we can measure their power and calculate the difference to determine the mode contrast $\Gamma = \beta_{\text{TE0}} - \beta_{\text{TE1}}$, which is used as a programming parameter. Fig. 2f demonstrates the multi-level programmability of the PMMC, in which $\Gamma$ is sequentially set to 64 distinguishable levels between -0.73 to +0.67 at 1555 nm. Since the theoretical range of $\Gamma$ is $(-1,1)$, it is an ideal parameter to represent the elements in the matrix *w*, with both positive and negative values, in multiply-accumulate (MAC) operation: *x* → *x* · *w* + *b*, where *b* is the bias parameter. MAC is the constitutional step of matrix-vector multiplication (MVM) in all neural network algorithms. The PMMC allows storing $w$ by programming $\Gamma$ in the GST metasurface as a nonvolatile memory. In-memory MAC computing can be performed with the PMMC by a measurement of the transmitted power when the input data $x$ is encoded in the power of the input optical signal. The lower inset of Fig. 2f shows the histograms of 20 repeated programming operations to set the PMMC mode contrast at two adjacent levels (levels 30 and 31), respectively. The well separated histograms clearly demonstrate the device's programming resolution and accuracy. The demonstrated 64-level programmability of the PMMC—the highest to the best of our knowledge for phase-change photonic devices[24]—corresponds to 6-bit resolution in setting $w$, which is critical to the training and inference precision of the neural network[46,47].

We harness the PMMC's high-precision programmability and in-memory computing capability to demonstrate an optical convolutional neural network (OCNN)[28-30,48]. A typical CNN consists of an input layer and an output layer, which are connected by multiple hidden layers in between. The hidden layers usually consist of a series of convolutional layers followed by pooling layers and fully connected layers at the end. We design a prototype optical CNN using a small network of PMMCs to implement patch-kernel matrix multiplication to compute convolution. Fig. 3a illustrates the operation principle of the OCNN for image processing, where an input grayscale image of dimensions $n \times n$ is convolved with a kernel of dimensions $k \times k$ to compute an activation map of dimension $(n–k+1) \times (n–k+1)$, assuming the convolution stride is 1. When operating the OCNN, we group the input image into $(n–k+1)^2$ patches (the shaded area in the upper panel of Fig. 3a) with the same dimensions as the convolution kernel, $k^2$. Each patch corresponds to the receptive field of an element in the activation map accordingly. Thus, a convolution operation requires $(n–k+1)^2 \times k^2$ MAC operations in total, which is a high load of computation and can most benefit from optical computing's speed and energy advantages.

To compute the convolution, $(n–k+1)^2$ patch matrices of the input image are optically fed into the photonic kernel sequentially while the kernel elements, that is, the PMMCs, are programmed to fixed values. At each timeframe of the computation, the corresponding patch matrix is reshaped into a single column of data with the length $k^2$. The data is input into the optical system in $k^2$ channels as sequences of incoherent optical pulses, whose power amplitude is controlled by a variable optical attenuator (VOA) to encode the value of each pixel value $X_{ij}$ in the greyscale image. The corresponding element $W_{ij}$ of the kernel matrix is programmed as the mode contrast $\Gamma$ of each PMMC. The resulting transmitted power of TE$_0$ and TE$_1$ modes are then summed incoherently using two photodetectors. Their difference is calculated electronically and used in post-processing steps. As a result, the output will correspond to a time series of patch-kernel MVM with the amplitude encoding the values of the computation results, which is the activation map of convolution. Since the modal contrast $\Gamma$ of our PMMCs can assume both positive and negative values, it can represent the kernel matrix elements without the need of an additional offset, which otherwise would take additional steps to set in each computation cycle.

Experimentally, we build a small-scale, four-channel system with four PMMCs to represent a 2×2 kernel matrix, as shown in the optical images in Fig. 3b. As a demonstration, we perform the convolution of a 256×256 8-bit grayscale image of a cameraman (Fig. 3c) to detect its edge features. As shown in Fig. 3b, the TE$_0$ mode output coming from all the PMMCs is combined using on-chip Y-junctions, while the TE$_1$ mode output power is combined off-chip because the same ports are used to program the PMMCs optically. Because combining four incoherent sources using Y-junctions will inherently reduce the power by a factor of 4, we rescale the measured TE$_0$ mode power by this factor when calculating the power differences between two modes. To detect vertical and horizontal edges, kernel matrices as in the right column of Fig 3d and e are used, and so are the PMMCs programmed. Take the vertical edge detection for example, the kernel is set to be $\begin{bmatrix} -1 & 1 \\ -1 & 1 \end{bmatrix}$ so to compute the discrete first-order derivative, $X_{i+1,j} + X_{i+1,j+1} - X_{i,j} - X_{i,j+1}$, where $i, j$ are the indices of the input image matrix. Each kernel element $W_{ij}$ is stored as the mode contrast value $\Gamma$ in the corresponding PMMC, with $W_{ij}$ =1(-1) corresponds to the fully aGST (cGST) phase (see Supplementary Information for a more detailed description about the operation procedure). The computed images after convolution without any post-processing are shown in the left column of Fig. 3d, e, for horizontal and vertical edge detection, respectively. The two images are then added to produce the right image in Fig. 3b, which highlights silhouettes of the objects with sharp edges such as the cameraman and the buildings in the original image, while suppressing smooth features such as the sky and the water. The optically computed edge detection image also agrees very well with the calculated result using conventional image processing algorithms (see Supplementary Information). This result verifies the capacity and fidelity of optical convolution performed with the PMMC-based photonic kernel, which is a prerequisite for an OCNN.

Beyond the convolution layer, the MAC computation performed with optical signals and the PMMC network can also be applied to the pooling (average pooling) and fully connected layers, where the PMMCs are used as weight banks instead, to realize a complete OCNN. In our experiment, we sequentially reuse the PMMC array in both convolution and fully connected layers to demonstrate an OCNN and perform proof-of-concept imaging recognition tasks of distinguishing handwritten numbers "1" and "2" from the MNIST database. Fig. 4a illustrates the architecture and processes of the OCNN. The 28×28 pixels, 8-bit grayscale images of number "1" or "2" are fed into the input layer as optical signals. The data is then convolved with two 2×2 photonic kernels $K_1$ and $K_2$ to generate two 27 × 27 images of activation maps. After adding a bias $b_1$ and applying the nonlinear ReLu function, the output images are sent to an average pooling layer with a subsampling factor of 27, which reduces the images a 2×1 vector. This vector is then fed into the fully connected layer with a 2×2 photonic weight bank $K_3$ programmed in the PMMC array, added with a bias $b_2$ and applied the standard sigmoid function. The final output is a vector that gives the identified class of the input image, that is, $[1 \quad 0]^T$ corresponds to the number "1" and $[0 \quad 1]^T$ corresponds to the number "2". In this OCNN, the MVM computations such as the convolution and the fully connected layers are all performed optically with the PMMCs, whereas bias and nonlinear functions are realized electronically.

Before using the OCNN, we first train all the parameters in the layers with the standard back-propagation algorithm using the gradient descent method[49]. The training set consists of 11000 images of the handwritten number "1" or "2" from MNIST training images (see supplementary for training details). The training yields values for each element in the convolutional kernels and the weight bank, as shown in Fig. 4b and d. We then program the PMMC array to represent these elements. In Fig. 4c, we show the raw data of the convolutional activation maps encoded in time series of optical signals, which is the output from the PMMC array after the input image convolves with the photonic kernel $K_1$ and $K_2$. Since each photonic processing layer results in electrical signals output from the photodetectors, electronic post-processing is performed to add bias and apply nonlinear function and pooling. The resultant data is re-coded into optical signals and fed to the next photonic layer. Further experimental details are included in the supplementary information. We evaluate the system's performance after training on a recognition test set, which consists of 100 randomly chosen "1" or "2" images (55 number "1" and 45 number "2") from the MNIST testing image database. Fig. 4e shows the result that our OCNN correctly identified 91 out of 100 cases (9% error rate), which compares squarely with the result of a computer (10% error rate). The slight difference is mainly caused by the small deviation of the experimentally programmed values in the matrices ($K_1$, $K_2$ and $K_3$) from the trained values, which occurs when the system's conditions drift during operation. This result successfully demonstrates the OCNN's viability and accuracy in performing standard neural network algorithms.

In summary, we have demonstrated a compact programmable waveguide mode converter using GST-based phase-gradient metasurface with high programming resolution, efficiency and broadband operation. We have built a photonic kernel based on an array of such PMMC devices and implemented an optical convolutional neural network to perform image processing and recognition tasks. Our results show that phase-change photonic devices, such as the PMMC demonstrated here, can enable robust and flexible programmability and realize a plethora of unique optical functionalities that are scalable for large-scale optical computing and neuromorphic photonics. Although optical computation in this work is performed at a low speed of ~1 kHz by using low-speed VOAs to encode data into optical signals, state-of-the-art integrated photonic transmitters and photodetectors can drive the system at a speed of many 10s of Gbits/sec[50,51]. Using wavelength division multiplexing (WDM) can further increase the number of parallel computation. The 2×2 array prototype system demonstrated in this work performs optical computation incoherently in a broadband. It can be readily scaled up toward a large network using a photonic crossbar array architecture[52-57] (See Supplementary Information for details of such a design), and compares favorably with other photonic computing schemes using coherent methods[30] or optical resonators[28,58,59]. The feasible size ($n \times m$) of such crossbar arrays will not be limited by the insertion loss of the PMMC (~7 dB for TE1 mode, Fig. 2d); rather it will be limited by the directional couplers with coupling efficiency of $1/n$, as is needed to equally combine signals from $n$ units. The PMMC device is very compact with a footprint of ~80×20 μm (Fig. 2a, including the mode selector). When performing MAC compuating with a high speed, an OCNN system using the device can afford an extremely high areal computing density (defined as MAC operations per time per unit area). For example, assuming a moderate datarate of 10 Gbits/sec and 4 WDM wavelengths in parallel per channel, the computing density will reach an upperbound value of 25 TOPS/mm$^2$ (Tera-Operations per second per mm$^2$), which is significantly higher than that of digital electronic accelerators such as GPUs[60] and tensor processing units (TPUs)[61,62]. Using silicon instead of silicon nitride can further reduce the device footprint to increase the computing density[63]. Besides MAC operation, the equally important computing processes of applying nonlinear functions and pooling can also be achieved optically by using elements such as nonlinear optical resonators, modulators, and amplifiers[27,28,64]. Alternatively, a hybrid photonic-electronic system may optimally balance energy-efficiency and speed advantages of photonic systems, while realizing flexible non-linearity, connectivity, and training precision using microelectronics[26,65,66]. With these advances, the photonic neural network accelerator will be very promising for AI in data centers where massive optical interconnects have already been deployed.

**Acknowledgment:**

We acknowledge the funding support provided by the ONR MURI (Award No. N00014-17-1-2661, Program Manager: Dr. Brian Bennett).

**Data Availability Statement**

The data that support the findings of this study are available from the corresponding author upon reasonable request.

**Code Availability Statement**

No custom computer code or mathematical algoritm is used to generate results that are reported in this study.

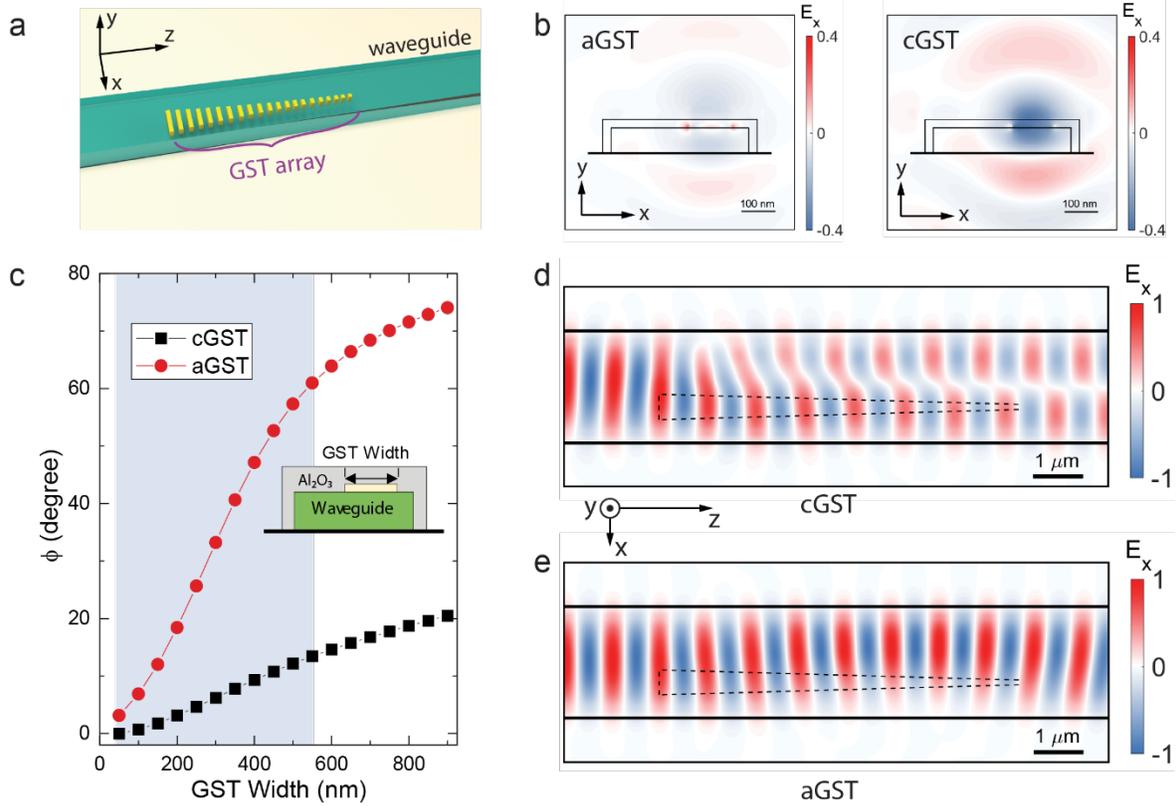

**Figure 1** Design of the phase-gradient metasurface mode converter. **a.** 3D illustration of the devices. **b.** Finite element simulation of the scattered electric field by one nano-antenna when the GST is in aGST (left panel) and cGST (right panel) phases, respectively, showing the distinctive difference. **c.** The phase of the scattered mode as a function of the GST nano-antenna width for cGST and aGST phases. The shaded region indicates the range of antenna widths that are used in the phase gradient metasurface. Inset: cross-sectional view of the structure. **d, e.** FDTD simulation results showing effective mode conversion from the $TE_0$ mode to the $TE_1$ mode when the GST is in crystalline phase (d), but only a small perturbation when the GST is in amorphous phase (e).

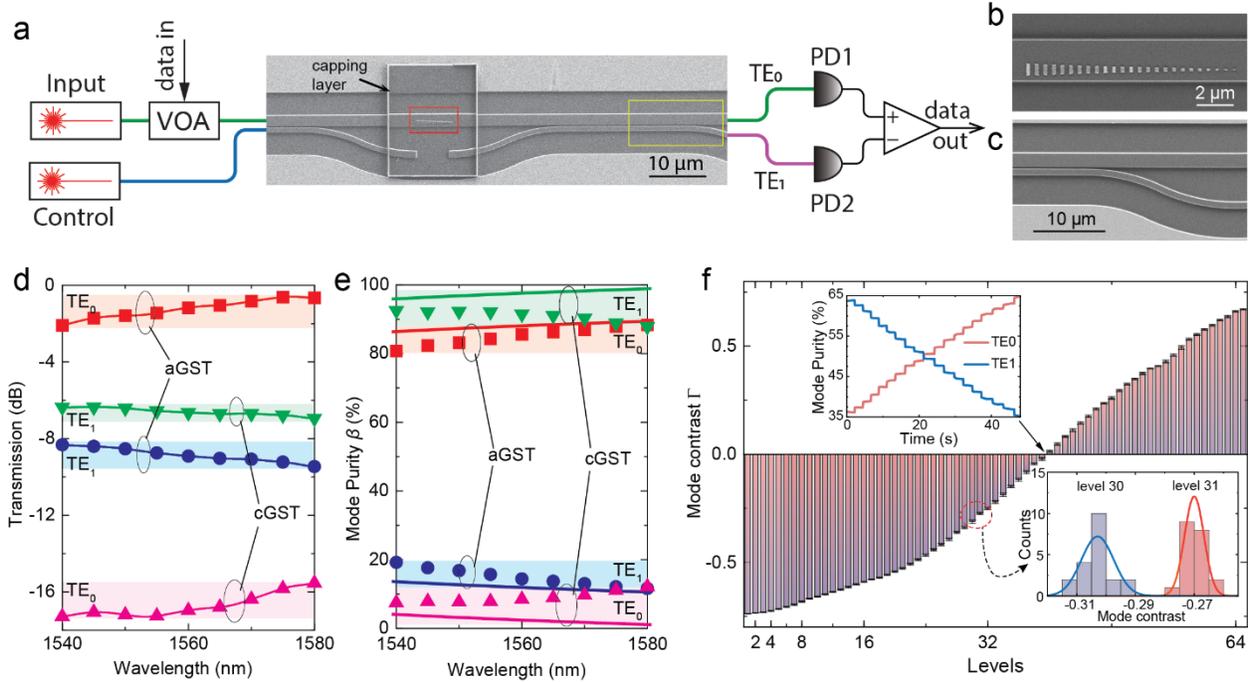

**Figure 2** Operation of the programmable metasurface mode converter (PMMC). **a.** Scanning electron microscope (SEM) image of the complete device and the measurement and control schematics. The complete PMMC device consists of an encapsulated GST phase gradient metasurface (red box) and a mode selectors (yellow box) and. The white box appears from the edge of 218 nm thick $Al_2O_3$ encapsulating layer. **b.** Zoomed-in SEM image of the phase-gradient metasurface on the waveguide before depositing the $Al_2O_3$ layer encapsulation for better imaging. **c.** Zoom-in SEM image of the $TE_0$/$TE_1$ mode selector. **d.** The transmission coefficient (insertion loss) of the devices for $TE_0$ and $TE_1$ modes and aGST and cGST phases. The transmission for the $TE_0$ mode is switched with a high extinction ratio >16 dB or 4000%. **e.** The mode purity is controlled by the mode converter to >80% for both modes. **f.** The programable mode converter controls the mode contrast $\Gamma$ at 64 distinct levels, corresponding 6-bit programming resolution. Upper inset: zoomed-in view of the contrast levels. Lower inset: histograms of 20 programming operations to set the contrast two adjacent levels (30 and 31). The well-separated histograms demonstrate the programming repeatability and accuracy.

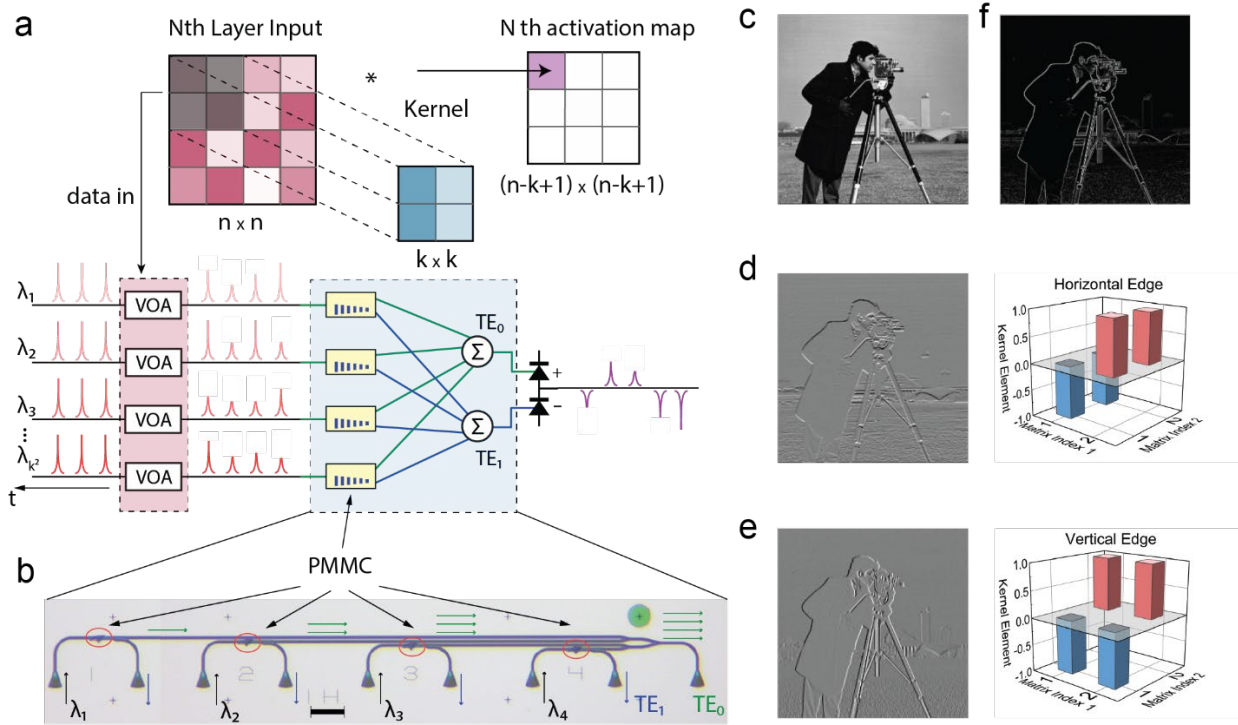

**Figure 3** Using a PMMC array as a photonic computing core for convolutional image processing. **a.** Schematic of optical convolution for image processing. An array of $k^2$ PMMC is programmed to store the kernel matrix. A patch of pixels of the image is encoded as optical pulses and input into $k^2$ optical channels to perform MAC operation with the kernel. The output in $TE_0$ and $TE_1$ are summed incoherently and measured with photodetectors. The activation map is represented by the mode contrast and could be both positive and negative. **b.** Optical microscope image of the photonic core consisting of four PMMCs with four input channels. The $TE_0$ mode outputs are summed on-chip with Y-junctions whereas $TE_1$ mode outputs are summed off-chip. Optical control pulses are input using the same set of grating couplers used for the $TE_1$ mode detection. **c.** A greyscale image of a cameraman used as the input image. **d.** and **e.** Left: the raw image generated by convolution with the kernel matrix for detection of horizontal (d) and vertical (e) edges. Right: the corresponding kernel matrix for edge detection. **f.** Combined image of horizontal and vertical edge detection, highlighting all the sharp edges in the original image.

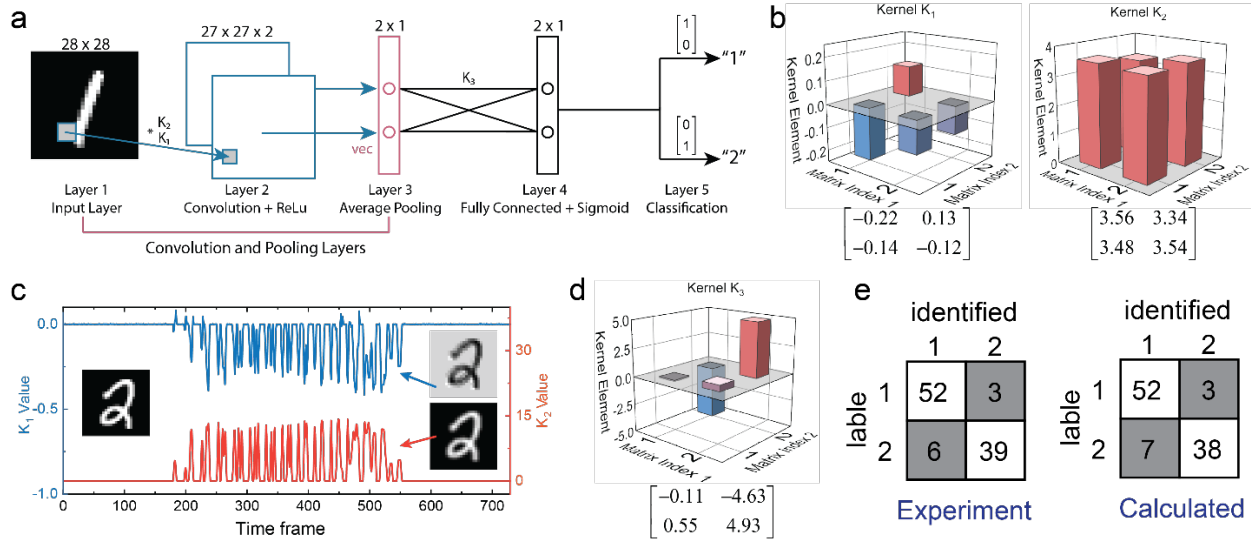

**Figure 4** Building an optical CNN for imaging recognition. **a.** Operation procedure of using the optical CNN to recognize handwriting numbers from the MNIST database. The optical CNN consists of a convolution layer with two kernels, a pooling and a fully connected layer. The output gives the answer whether the input image is "1" or "2". **b.** The convolution kernel matrices $K_1$ and $K_2$ generated by training the CNN. **c.** Raw output data of the convolution layer of two kernel matrix. **d.** The weight bank matrix used in the fully-connected layer. **e**. The recognition results from the experiment with the optical CNN (left) and calculation with a computer (right) show an excellent agreement.